\documentclass[aps,prd,reprint,preprintnumbers,nofootinbib]{revtex4-1}
\pdfoutput=1

\usepackage{xcolor}
\usepackage{graphicx}
\usepackage{amsmath}
\usepackage{amssymb}
\usepackage{soul}

\usepackage[colorlinks]{hyperref}
\hypersetup{allcolors=[rgb]{0,0,1}}

\newcommand{\orcid}[1]{\href{https://orcid.org/#1}{#1}}

\usepackage{scalerel}
\newlength{\heightnu}
\AtBeginDocument{\settoheight{\heightnu}{$\nu$}}

\usepackage[normalem]{ulem}

\definecolor{ablue}{RGB}{51, 173, 255}

\newcommand{\mn}{\Delta m_{21}^2}

\def\lsim{\raise0.3ex\hbox{$\;<$\kern-0.75em\raise-1.1ex
\hbox{$\sim\;$}}}
\def\gsim{\raise0.3ex\hbox{$\;>$\kern-0.75em\raise-1.1ex
\hbox{$\sim\;$}}}

\keywords{Neutrino Physics, Reactor Experiments}

\begin{document}

\vspace*{-3cm}
\title{ 
Constraint on the solar $\Delta m^2$ using 4,000 days of short baseline reactor neutrino data
}

\author{Alvaro Hernandez Cabezudo}
\email{alvaro.cabezudo@kit.edu}
\thanks{\orcid{orcid \# 0000-0001-9594-5450}}
\affiliation{Institut f\"{u}r Kernphysik, Karlsruher Institut f\"{u}r Technologie (KIT), D-76021 Karlsruhe, Germany}

\author{Stephen~J.~Parke}
\email{parke@fnal.gov} 
\thanks{\orcid{orcid \# 0000-0003-2028-6782}}
\affiliation{Theoretical Physics Department, Fermi National Accelerator Laboratory, Batavia, IL 60510, USA }

\author{Seon-Hee~Seo}
\email{sunny.seo@ibs.re.kr}
\thanks{\orcid{orcid \# 0000-0002-1496-624X}, corresponding author}
\affiliation{
Center for Underground Physics, Institute for Basic Science, 
 Daejeon 34126, Korea}

\preprint{FERMILAB-PUB-19-190-T}

\vglue 2.6cm

\vspace*{3.cm}

\begin{abstract}
There is a well known 2$\sigma$ tension in the measurements of the solar $\Delta m^2$ between KamLAND and SNO/Super-KamioKANDE. 
Precise determination of the solar $\Delta m^2$ is especially important in connection with current and future long baseline CP violation measurements. 
Reference~\cite{Seo:2018rrb} points out that currently running short baseline reactor neutrino experiments, Daya Bay and RENO, 
can also constrain solar $\Delta m^2$ value as demonstrated by a GLoBES simulation with a limited systematic uncertainty consideration. 
In this work, the publicly available data, from Daya Bay (1,958 days) and RENO (2,200 days) are used to constrain the solar $\Delta m^2$. 
Verification of our method through $\Delta m^2_{ee}$ and $\sin^2 \theta_{13}$ measurements is discussed in Appendix~\ref{a:validation}.
Using this verified method, reasonable constraints on the solar $\Delta m^2$ are obtained using above Daya Bay and RENO data, both individually and combined. We find that the combined data of Daya Bay and RENO set an upper limit on the solar $\Delta m^2$ of 18 $\times 10^{-5}$ eV$^2$ at the 95\% C.L., including both systematic and statistical  uncertainties. This constraint is slightly more than twice the KamLAND value. 
As this combined result is still statistics limited, even though driven by Daya Bay data, the constraint will improve with the additional running of this experiment. \\
\end{abstract}

\vspace*{1.cm}

\date{July 8, 2019}


\maketitle

\section{Introduction}

Evidence that neutrinos are massive and mix is well established by a significant number of experiments. In this paper, we are interested in the mass squared difference, $\Delta m^2_{21}$; the mass squared difference of the two mass eigenstates that have the greatest fraction of electron neutrino, $\nu_1$ and $\nu_2$.  This mass splitting is responsible  for the neutrino flavor transformations that occur inside the Sun (hence the name the solar mass squared difference), and for the antineutrino oscillations observed at an L/E $\sim$ 15~km/MeV.

In this paper, we use publicly available data to follow up a recent paper  \cite{Seo:2018rrb}, that Daya Bay  \cite{An:2015rpe} and RENO \cite{RENO:2015ksa}, the short baseline  ($\sim$1.5~km)  reactor antineutrino experiments currently running , have enough data already collected to constrain  $\Delta m^2_{21}$.

The combined constraint by Daya Bay and RENO, gives an important consistency check of the standard three neutrino paradigm as well as adding addition information to the size of $\Delta m^2_{21}$.
The  $\sim$2$\sigma$ tension between the combined Super-Kamiokande (SK)~\cite{Abe:2010hy} \& Sudbury Neutrino Observatory (SNO)~\cite{Aharmim:2011vm} solar neutrino 
measurements and KamLAND~\cite{Gando:2010aa} reactor experiment ($L/E \sim$ 50~km/MeV) is not directly addressed by this constraint. 
However such a combined Daya Bay plus RENO constraint is at a different $L/E$ range ($\sim$ 0.5 km/MeV) than the above mentioned measurements 
as well as JUNO~\cite{An:2015jdp}. 
Moreover, the ratio of  $\Delta m^2_{21}$  to  $\Delta m^2_{31}$, at an  $L/E \sim$ 0.5 km/MeV, is required for the precision measurement of leptonic CP violation parameter, 
by NOvA~\cite{Ayres:2004js}, T2K~\cite{Abe:2011ks} and future Long Baseline (LBL) experiments. 

Currently there are two measurements of the solar mass squared difference, $\Delta m^2_{21}$.
One measurement comes from a combined measurement by SNO and SK using the the observation of a day-night asymmetry by SK and the non-observation 
of the low energy up turn of the $^8$B neutrino survival probability by SNO and SK. This combined result is 
\begin{eqnarray}
\Delta m^2_{21}  = 5.1 ^{\,+1.3}_{\,-1.0}  \times 10^{-5} ~{\rm eV}^2,
\label{eq:sk+sno}
\end{eqnarray}
from SNO and SK. Similar results are obtained by Nu-Fit \cite{Esteban:2018azc}.
The other measurement is from KamLAND, the long baseline reactor anti-neutrino experiment, see \cite{Gando:2010aa}, at
\begin{eqnarray}
\Delta m^2_{21}  = 7.50^{\,+0.20}_{\,-0.20} \times 10^{-5} ~{\rm eV}^2,
\label{eq:kamland}
\end{eqnarray}
If CPT invariance is a good symmetry of nature then  the $\mn$ measured from solar neutrinos  and reactor anti-neutrinos is required to give the same value. Currently this important parameter for neutrino physics suffers from a 2$\sigma$ level tension. This tension could come from  new physics,  some error in the analysis of one or more of the experiments or a statistical fluctuation.

Moreover, the ratio of  $\Delta m^ 2_{21}$ to $\Delta m^ 2_{31}$ is required for the determination of the CP phase, $\delta$, in the  long baseline neutrino\footnote{In the rest of this paper, when referring to neutrinos, we mean both neutrinos and/or anti-neutrinos.} oscillation experiments (NOvA, DUNE \cite{Acciarri:2015uup}, T2K, T2HK \cite{Abe:2015zbg}, T2HKK \cite{Abe:2016ero}) as the size  of the CP violation is proportional to $\Delta m^ 2_{21}$ to $\Delta m^ 2_{31}$, as well as the Jarlskog invariant. At $L/E \sim$ 500 km/GeV=0.5 km/MeV, the first oscillation peak in vacuum, for $\nu_\mu \rightarrow \nu_e$
\begin{eqnarray}
 P(\bar{\nu}_\mu \rightarrow \bar{\nu}_e)-P(\nu_\mu \rightarrow \nu_e) \, \approx \, \pi \, J \, \left( \frac{\Delta m^2_{21}}{\Delta m^2_{31}}\right)
\end{eqnarray}
where the Jarlskog invariant, J, is\\
 $J  =  \sin 2 \theta_{12} \sin 2 \theta_{13} \cos \theta_{13} \sin 2 \theta_{23} \sin \delta \, \approx 0.3  \sin \delta $.
  
In the bi-event plane for T2K, see Fig 44 of \cite{Abe:2017vif}, 
\begin{eqnarray}
N(\nu_\mu \rightarrow \nu_e)=37 \quad {\rm and} \quad N(\bar{\nu}_\mu \rightarrow \bar{\nu}_e)  = 4  \nonumber
\end{eqnarray}
is outside the allowed region (by about 1 $\sigma$). This can be well accommodated  by  a $\Delta m^2_{21}$ value, approximately twice the  KamLAND value.
Again, this is probably a statistical fluctuation but with only the KamLAND precision measurement of $\Delta m^2_{21}$, other possibilities are still viable.

The future medium baseline, $L/E \sim$ 15 km/MeV,  reactor experiment JUNO will measure to better than 1\% precision  $\Delta m^2_{21}$ and $\sin^2 \theta_{12}$, 
see~\cite{An:2015jdp}. JUNO experiment is currently under construction and their precision measurements of $\Delta m^2_{21}$ and $\sin^2 \theta_{12}$ will not be available until approximately 5 years from now.  
Later next decade, the proposed experiments Hyper-K \& DUNE will also give us precision measurements of $\Delta m^2_{21}$ using $^8$B solar neutrinos, see  \cite{Abe:2018uyc}  and  \cite{Beacom:2018xyz} respectively.

In section II, we briefly discuss in detail the effects of increasing $\mn$ on the $\bar{\nu}_e$ survival  probability.   
Then in section III 
Daya Bay and RENO data sets used in this work are discussed followed by section IV, V, and VI for methods and systematic uncertainties, 
results, and conclusion, respectively. 
In Appendix~\ref{a:validation} it is described the verification of the method used in this work by comparing $\Delta m^2_{ee}$ vs $\sin^2 2 \theta_{13}$ measurements. 
In Appendix~\ref{a:predEvt} we describe expected events and how pull parameters are inserted. 
In Appendix~\ref{a:fixedVsFree} the effects of fixing or floating the value  of $\Delta m^2_{ee}$ are discussed.

\section{Survival Probability}

In vacuum, the electron antineutrino survival probability is
\begin{eqnarray}
& & P(\bar{\nu}_e \rightarrow \bar{\nu}_e )  =  1-P_{12}-P_{13} \quad {\rm with}  \\[2mm]
P_{12} & = & \sin^2 2 \theta_{12} \cos^4\theta_{13}  \sin^2 \Delta_{21},  \nonumber \\[2mm]
P_{13} & = & \sin^2 2 \theta_{13}\, ( \cos^2 \theta_{12} \sin^2 \Delta_{31}+\sin^2 \theta_{12} \sin^2 \Delta_{32}),  \nonumber \end{eqnarray}
where the kinematic phases are given by $\Delta_{jk} \equiv \Delta m^2_{jk} L/(4E)$ and $\theta_{13} \approx 8^\circ $ and  $\theta_{12}\approx 33^\circ $ are the reactor and solar mixing angles respectively. 
 The $P_{12}$ term is associated with the solar oscillation scale of 15 km/MeV and the $P_{13}$ term  is associated with the atmospheric oscillation scale of 0.5 km/MeV.  To excellent fractional precision\footnote{The fractional precision is better than 0.05\% for L/E $<$ 1 km/MeV. Also, in this L/E range, the exact $P_{13}$ is very insensitive to mass ordering provided the value of $|\Delta m^2_{ee}|$ is the same for both mass orderings. }, the $P_{13}$ term can be approximated by
\begin{eqnarray}
P_{13} \approx \sin^2 2 \theta_{13} \sin^2 \Delta_{ee} 
\end{eqnarray}
where $\Delta m^2_{ee} \equiv \cos^2 \theta_{12} \Delta m^2_{31}+ \sin^2 \theta_{12} \Delta m^2_{32}$
\cite{Nunokawa:2005nx,Parke:2016joa},  interpreted as the $\nu_e$ average of $\Delta m^2_{31}$ and $\Delta m^2_{32}$.

Using the fit values given in \cite{Esteban:2018azc}, and an $L/E$ range  around the first oscillation minimum  ($L/E \sim 0.5\,{\rm km/MeV} $),  $P_{12}$ and $P_{13}$ is well approximated by:
\begin{eqnarray}
P_{12} & \approx & 0.002  \left(\frac{L/E}{0.5 \,{\rm km/MeV}}\right)^2  \left(\frac{\Delta m^2_{21}}{7.5 \times 10^{-5} \,{\rm eV^2}}\right)^2
  \\[2mm]
P_{13} & \approx & 0.08 \sin^2 \left( \frac{\pi}{2} \left(\frac{L/E}{0.5 \,{\rm km/MeV}}\right) \right).
 \end{eqnarray}
 The $P_{12}$ term is almost negligible for all  $L/E < 1\,{\rm km/MeV}$, if $ \Delta m^2_{21}= 7.5 \times 10^{-5} \,{\rm eV^2}$.
For Daya Bay and RENO this covers the full $L/E$ range.

\begin{table*}[t]
\begin{center}
\caption{\label{t:events}
Live days (not operational days), effective baseline distance ($L_{\rm eff}$), observed IBD and background events for Daya Bay and RENO used in this work. 
For Daya Bay there are two near detectors ($N1$ and $N2$) in different sites.\\
}
\begin{tabular*}{0.8\textwidth}{@{\extracolsep{\fill}} c c c c}
\hline
\hline
                   &         & Daya Bay   & RENO \\
\hline
 Live days & Near ($N1 , N2$) & (1,637.12 , 1,647.64) &  1,807.88 \\
      & Far   & 1,692.69             &  2,193.04  \\
\hline
$L_{\rm eff}$ (m) & Near ($N1 , N2$)  &  (562.2 , 594.2)  & 430.4 \\
                   & Far     &  1637  &  1445.4 \\
\hline
Total \# of IBD events & Near ($N1 , N2$) &  (1,763,939 , 1,651,088)  &  833,433  \\
                & Far    &   486,873  &  98,292   \\
\hline
Total \# of background events  &  Near ($N1 , N2$) & (19,056 , 13,634)  &  17,229 \\
                         &   Far   &  2,230 & 4,912 \\
\hline
\hline
\end{tabular*}
\end{center}
\end{table*}

Suppose that  $ \Delta m^2_{21}$ is 3 times larger than KamLAND value, i.e. $ 22.5 \times 10^{-5} \,{\rm eV^2}$, then 
\begin{eqnarray}
P_{12} & \approx & 0.02  \left(\frac{L/E}{0.5 \,{\rm km/MeV}}\right)^2  \left(\frac{\Delta m^2_{21}}{22.5 \times 10^{-5} \,{\rm eV^2}}\right)^2 .
\end{eqnarray}
Now $P_{12}$ is now no longer tiny compared to  $P_{13}$ at $L/E =0.5\,{\rm km/MeV}$, oscillation minimum, and as  $L/E$ gets larger than 0.5 km/MeV, $P_{12}$ gets bigger, whereas $P_{13}$ is getting smaller. At an  $L/E = 1\,{\rm km/MeV}$, $P_{12}$ would be approximately equal to $\sin^2 2 \theta_{13}$ (0.08) for this value of  $ \Delta m^2_{21}$.
It is this quadratic rise in $P_{12}$  as  $ \Delta m^2_{21}$ increases that we exploit to place an upper limit on $ \Delta m^2_{21}$.
For further details on the survival probability as $ \Delta m^2_{21}$ increases see \cite{Seo:2018rrb}.

\section{\label{sec:simulation} Daya Bay and RENO Data Sets}

In this work, 1,958 days of Daya Bay data \cite{Adey:2018zwh} and 2,200 days of RENO data \cite{Bak:2018ydk} are used, 
where Daya Bay has about five times more inverse beta decay (IBD) events than RENO in their far detectors. 
Daya Bay data including background estimation, energy response function, and systematic uncertainties are taken from the supplementary material in \cite{Adey:2018zwh}. 
RENO data and background estimation are extracted from FIG.1 in \cite{Bak:2018ydk} and systematic uncertainties are also taken from \cite{Bak:2018ydk}.
Table~\ref{t:events} shows summary of the basic parameters, i.e., $L_{\rm eff}$, IBD rate, and background rate, 
for near and far detectors of Daya Bay and RENO used in this analysis. 
Note that there are two near detectors in different sites for Daya Bay.

\section{Methods and Systematic Uncertainties}

Best fit values on $\Delta m^ 2_{21}$ and $\sin^2 2\theta_{13}$ are obtained by finding minimum $\chi^{2}$ values between data and predictions 
for all possible combination of the two parameters. 
Far-to-near ratio method is employed in this $\chi^2$ analysis to avoid the spectral shape anomaly around 5~MeV region~\cite{Seo:2014xei}
as well as to reduce systematic uncertainties.

The $\chi^{2}$ formalism as written below contains a covariance matrix (V$_{{\rm stat,}ij}$) to include statistical uncertainty and 
pull parameters ($\xi_{\alpha}$) to include systematic uncertainties. 
\begin{eqnarray*}
\chi^2 &=& \sum^{\rm N_{bins}}_{i,j} \left( D^{F/N}_i - P^{F/N}_i \right) V^{-1}_{{\rm stat,}ij} \left( D^{F/N}_j - P^{F/N}_j \right) \\
&& + \sum_{\alpha}^{\rm N_{pull}} \frac{(\xi_\alpha - 1)^2}{\sigma^2_\alpha} \, ,
\end{eqnarray*}

where, $D^{F/N}_i \equiv \frac{O^F_i-B^F_i}{O^N_i-B^N_i}$, $ P^{F/N}_i \equiv \frac{X^F_i}{X^N_i}$,  
and $F (N)$ and $i$ ($j$) represent the Far (Near) detector and $i^{th}$ ($j^{th}$)
prompt energy bin, respectively. 
Being $O$ the observed number of IBD candidate events, $B$ the estimated background number of events and $X$ the expected number of events for a given 
$\Delta m^ 2_{21}$ and $\sin^2 2\theta_{13}$ pair.
A total of 26 energy bins ($N_{bins}$) is used for RENO from 1.2 to 8.4~MeV. 
The same number of energy bins are used for Daya Bay from 0.7 to 12~MeV 
but two near detectors are taken into account in the $\chi^{2}$ formalism by replacing $N_{bins}$ to $2N_{bins}$
where for $1\leqslant i \leqslant \rm N_{bins}$, $F=$EH3 and $N=$EH1, and for ${\rm N_{bins}}+1 \leqslant i \leqslant 2{\rm N_{bins}}$, $F=$EH3 and $N=$EH2.

For both Daya Bay and RENO, systematic uncertainties on the relative detection efficiency, relative energy scale and the main background contributions 
are taken into account as summarized in Table \ref{table:systematics}.

\begin{table}[h]
  \centering
  \begin{tabular}{|l|c|c|}
  \hline 
  & Daya Bay & RENO \\
  \hline  
 \textbf{Source} & \multicolumn{2}{c|}{\textbf{Uncertainty \%}} \\
 \hline
 Detection efficiency & 0.13 & 0.21\\ 
 Energy scale & 0.2 & 0.15\\
 Li-He background & 30 & 5-8\\
 Fast neutron background & 13-17 & --\\
 Accidental background & 1 & --\\
  \hline  
  \end{tabular}
  \caption{Relative systematic uncertainties used in this work for Daya Bay and RENO, taken from \cite{Adey:2018zwh,Bak:2018ydk} respectively.}
  \label{table:systematics}
\end{table}

\begin{figure*}[t]
\centering
\includegraphics[width=0.4\textwidth]{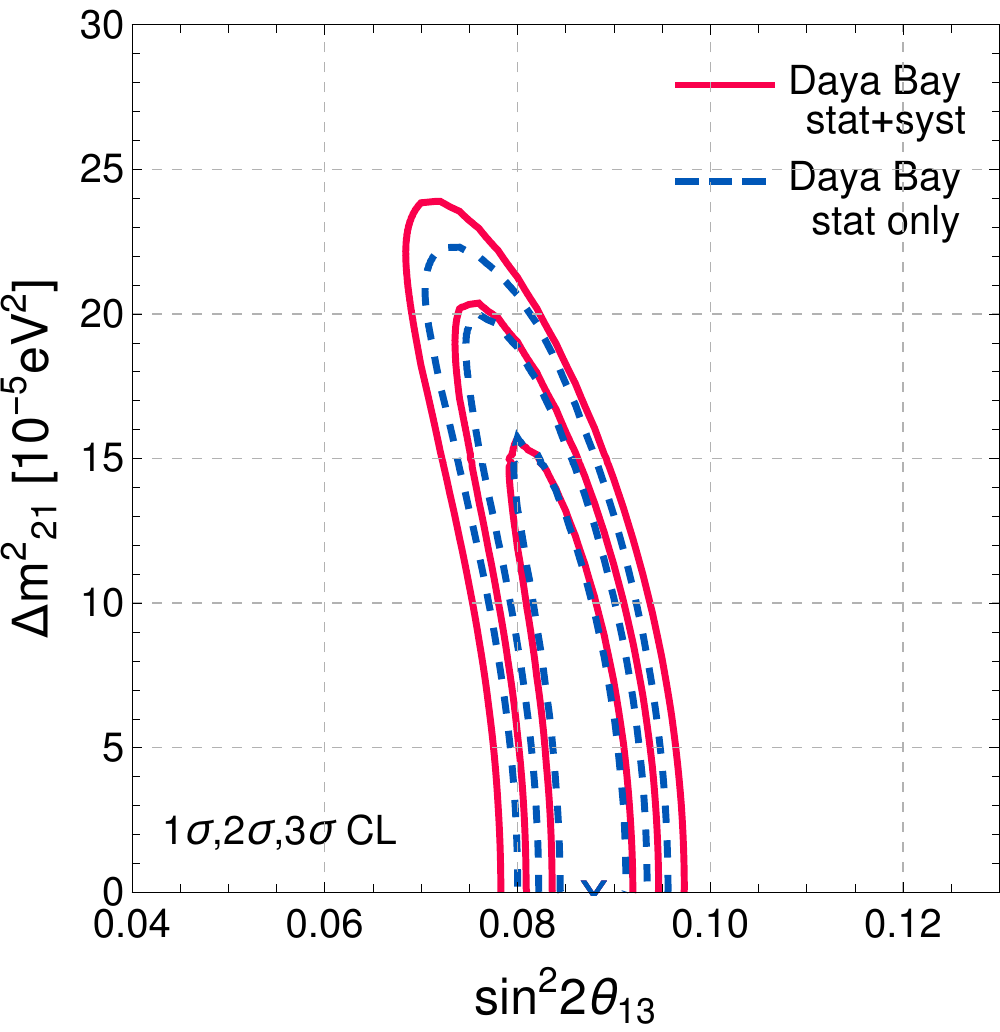}
\includegraphics[width=0.4\textwidth]{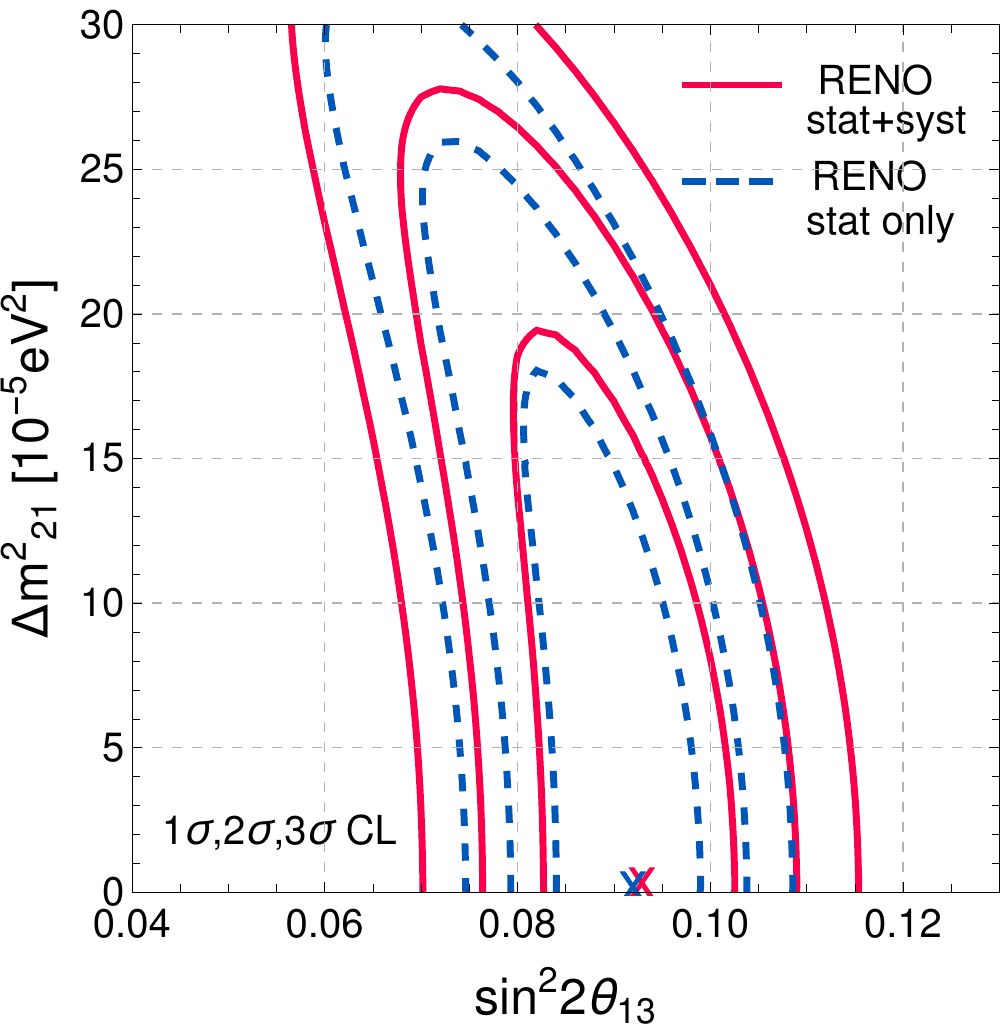}
\caption{The 1, 2 and 3 $\sigma$ allowed regions in the $\Delta m^2_{21}$ vs $\sin^2 2 \theta_{13}$ parameter space
for Daya Bay (1,958 days, 487 K IBD events at Far) (left panel) and  RENO (2,200 days, 98 K IBD events at Far) (right  panel).
In this fitting, the $\Delta m^{2}_{ee}$ value is constrained using the value from current long baseline (LBL) neutrino experiments,  see eq. \ref{eqn:eeprior}.
Red solid lines include both statistical and systematic uncertainties and blue dashed lines do only include statistical uncertainties.
The best fit values are shown with $\times$ signs.}
\label{fig:DB&RENO}
\end{figure*}

Besides the systematic uncertainties, additional systematic paddings (fudge factors) are added in our work to match Daya Bay and RENO results 
on $\theta_{13}$ and $\Delta m^{2}_{ee}$ measurements. 
For Daya Bay a 1.3 fudge factor to the relative energy scale and Li-He background uncertainties is added.
Whereas in RENO a 1.4 fudge factor is added to the relative detection efficiency uncertainty.
More details on the validation of our method and expected event description can be found in Appendices~\ref{a:validation} and \ref{a:predEvt}.
The RENO predictions are computed using the Daya Bay detector response function and the relative far-to-near normalization is computed 
comparing our total number of expected events with the total number of expected events in the RENO Far detector.
In order to match the best fit values of $\theta_{13}$ and $\Delta m^{2}_{ee}$ a 0.984 fudge factor is added to this normalization of a total event rate for RENO.
\\

\section{Results}

A 2-dimensional scan over $\Delta m^2_{21}$ and $\sin^2 2 \theta_{13}$ is performed to find the best fit value pair at the minimum value of $\chi^{2}$ described earlier, 
where in the oscillation probability, the parameter $\theta_{12}$ is fixed\footnote{A discussion on the effects of  varying $\theta_{12}$ in this analysis can be found 
in \cite{Seo:2018rrb}. } at $\sin^2 \theta_{12} = 0.310$.
The $\Delta m^2_{ee}$ parameter is constrained with a pull parameter, allowing it to vary within a $2\sigma$ range of a prior $\Delta m^2_{ee}$ value with a penalizing term
\begin{equation*}
\left( \frac{\Delta m^2_{ee {\rm ,\ prior}} - \Delta m^2_{ee}}{\sigma} \right)^2
\end{equation*}

The prior $\Delta m^2_{ee}$ value and its uncertainty are taken to be
\begin{equation}
\Delta m^2_{ee} = 2.45 \pm 0.15 \times 10^{-3} {\rm eV}^2
\label{eqn:eeprior}
\end{equation}
which is inferred from the combined measurement on $\Delta m^2_{\mu \mu}$ by current long baseline neutrino experiments in \cite{Esteban:2018azc}
through $\Delta m^2_{ee} \simeq \Delta m^2_{\mu\mu} \pm \cos 2 \theta_{12} \Delta m^2_{21}$, see \cite{Nunokawa:2005nx}, 
where the $+/-$ comes from the unknown mass ordering (NO/IO) and ignoring terms proportional to $\sin \theta_{13} \Delta m^2_{21}$. 
The unknown mass ordering is treated as an additional uncertainty (4\%) to $\Delta m^2_{\mu\mu}$ uncertainty (4\%) for the $\Delta m^2_{ee}$ uncertainty
which, therefore, becomes about 6\%. 

The best fit, 1, 2, and 3 $\sigma$ allowed regions of $\Delta m^2_{21}$ vs $\sin^2 2 \theta_{13}$ are shown in Fig.~\ref{fig:DB&RENO} 
with (solid lines) and without (dashed lines) systematic uncertainties for Daya Bay and RENO. 
Daya Bay result is better than RENO due to about five time more statistics. 

\begin{figure*}[t]
\centering
\includegraphics[width=0.4\textwidth]{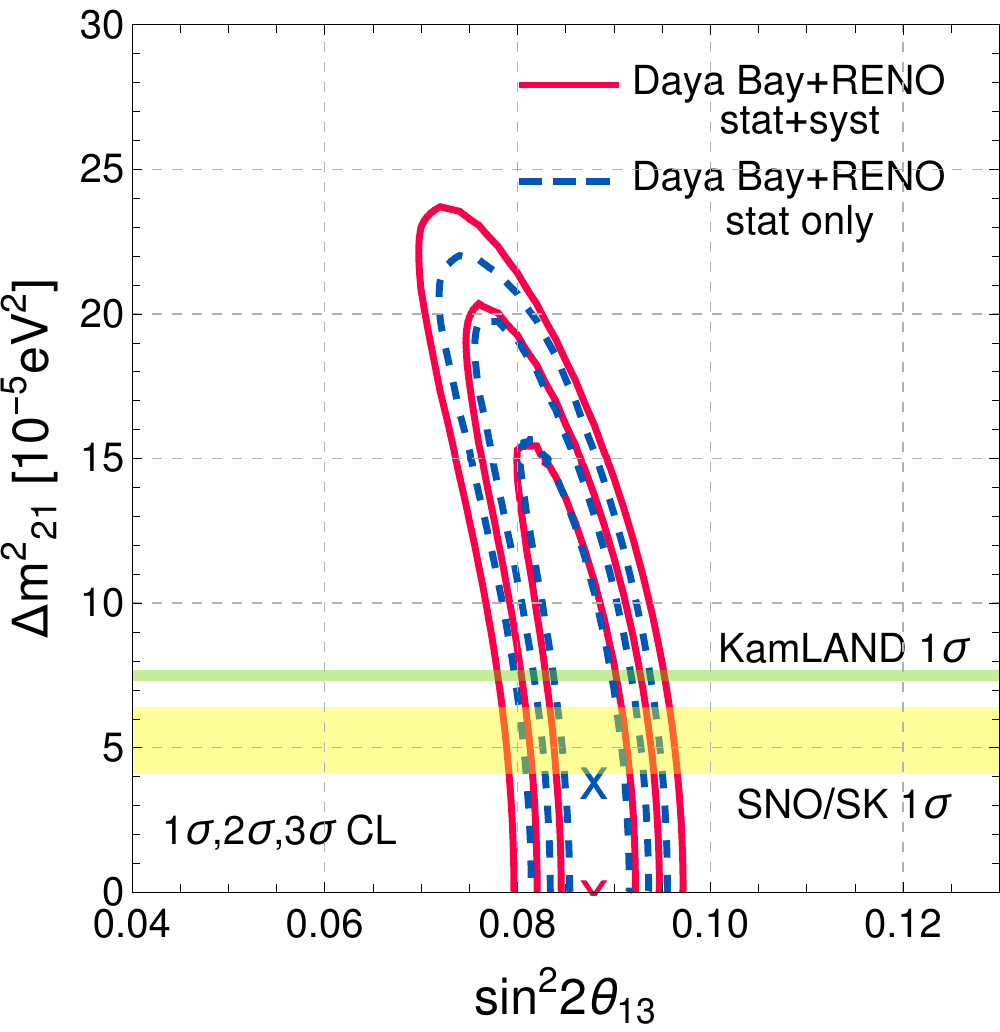}
\includegraphics[width=0.4\textwidth]{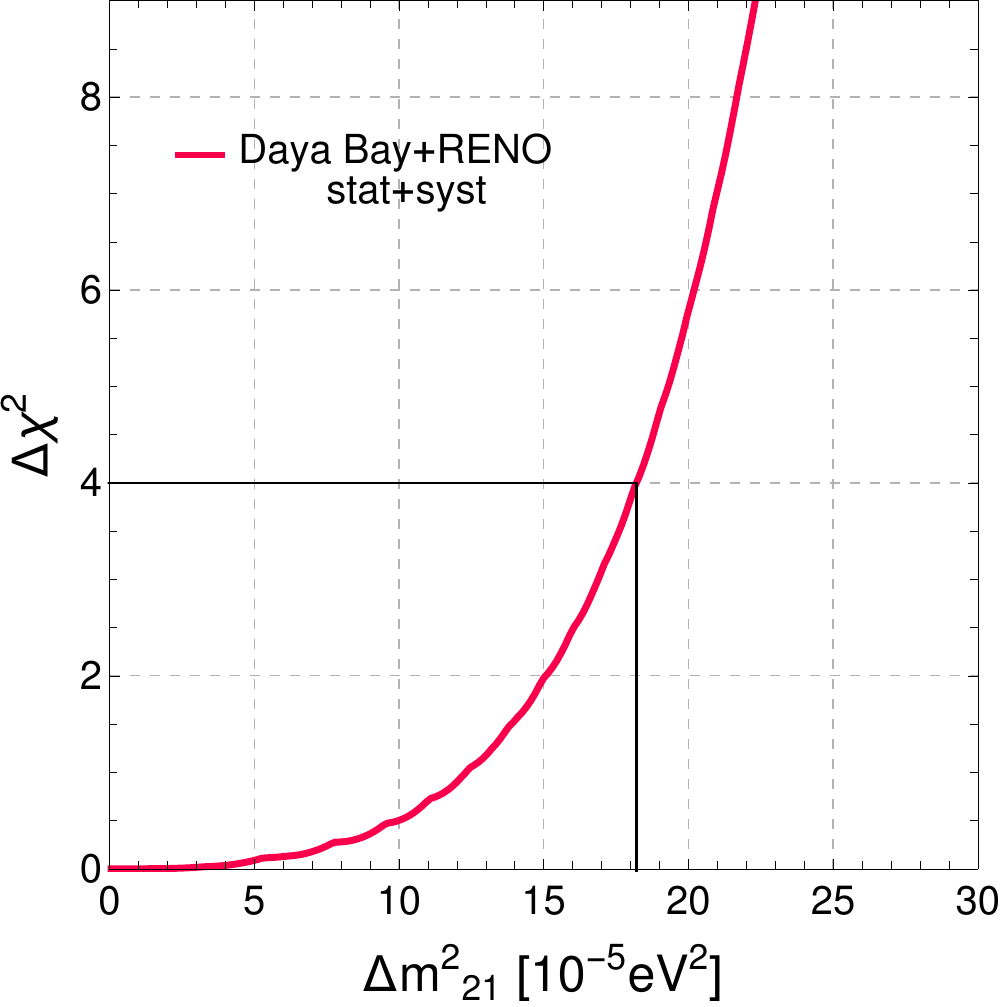}
\caption{Results of Daya Bay and RENO combined analysis. Left: 1, 2 and 3 $\sigma$  allowed regions in the $\Delta m^2_{21}$ vs $\sin^2 2 \theta_{13}$ parameter space.
Red solid lines include both statistical and systematic uncertainties and blue dashed lines only include statistical uncertainties.
The best fit values are shown with $\times$ signs. In this fitting, the $\Delta m^{2}_{ee}$ value is constrained using the value from current long baseline (LBL) neutrino experiments, see eq. \ref{eqn:eeprior}.
KamLAND and SNO/SK 1 $\sigma$ bands are overlaid for comparison.  This contour plot  shows that this measurement is still statistics limited.
Right: $\Delta \chi^2$ projection for $\Delta m^2_{21}$ measurement including systematic uncertainty, minimizing over $\sin^2 2 \theta_{13}$.
At the 2 $\sigma$ C.L. $\Delta m^2_{21}$ is constrained to be less than 18.3 $\times 10^{-5}$ eV$^2$.
(Note that in the abstract and conclusion we quote this result as 18 $\times 10^{-5}$ eV$^2$ at the 95\% C.L.)  }
\label{fig:comb}
\end{figure*}

To obtain the best result on solar $\Delta m^2$ measurement, 
a combined analysis of both Daya Bay and RENO data sets is performed. 
Figure~\ref{fig:comb} left plot shows the best fit, 1, 2, and 3 $\sigma$ allowed regions of $\Delta m^2_{21}$ vs $\sin^2 2 \theta_{13}$ of the combined analysis, 
and as expected the result is slightly improved by combining the two data sets. 
Figure~\ref{fig:comb} right plot shows the $\chi^2$ projection over $\Delta m^2_{21}$, obtained by minimizing over $\sin^2 2 \theta_{13}$.
 The upper bounds on $\Delta m^2_{21}$, including systematic uncertainties, are 12.3, 18.3 and 22.3 $\times 10^{-5}$ eV$^2$ at 1, 2 and 3 $\sigma$ CL, respectively.  
Current upper bounds are limited by statistics. 

\begin{figure*}[t]
\centering
\includegraphics[width=0.3\textwidth]{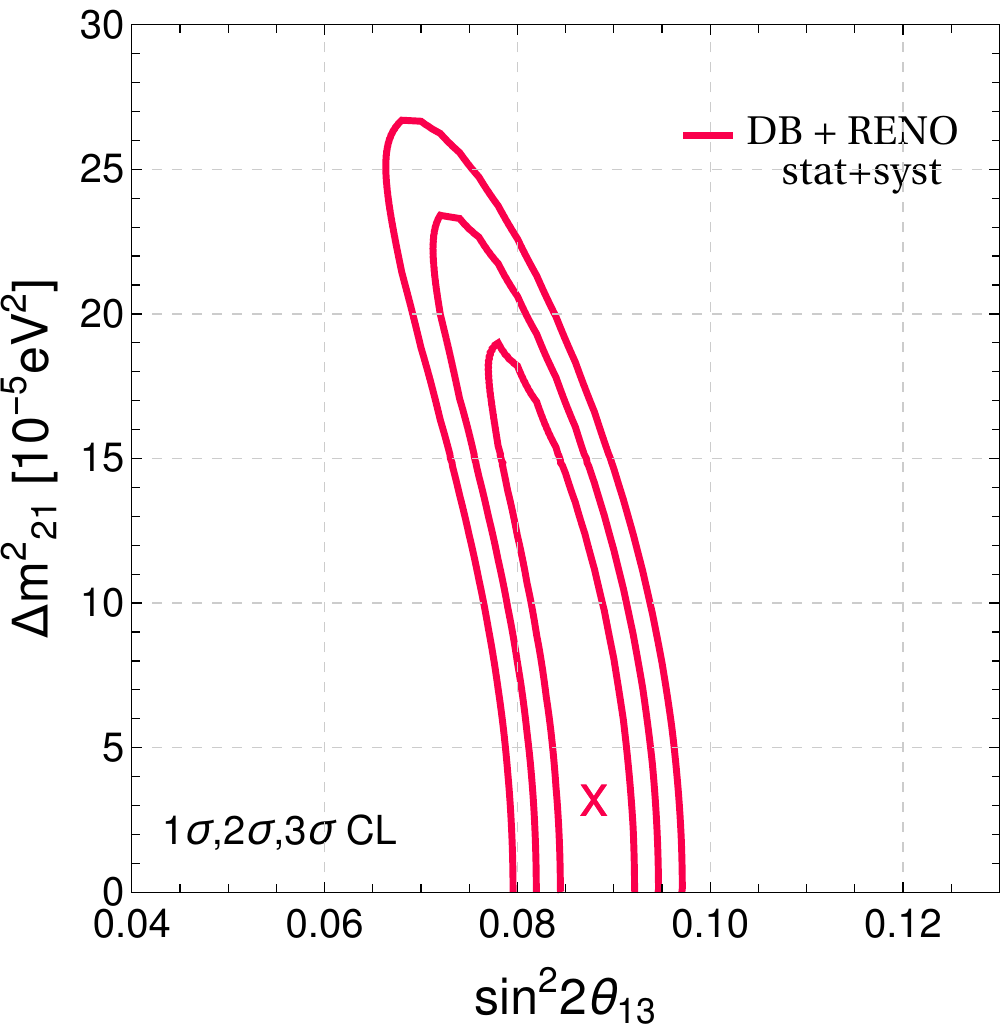}\hfill
\includegraphics[width=0.3\textwidth]{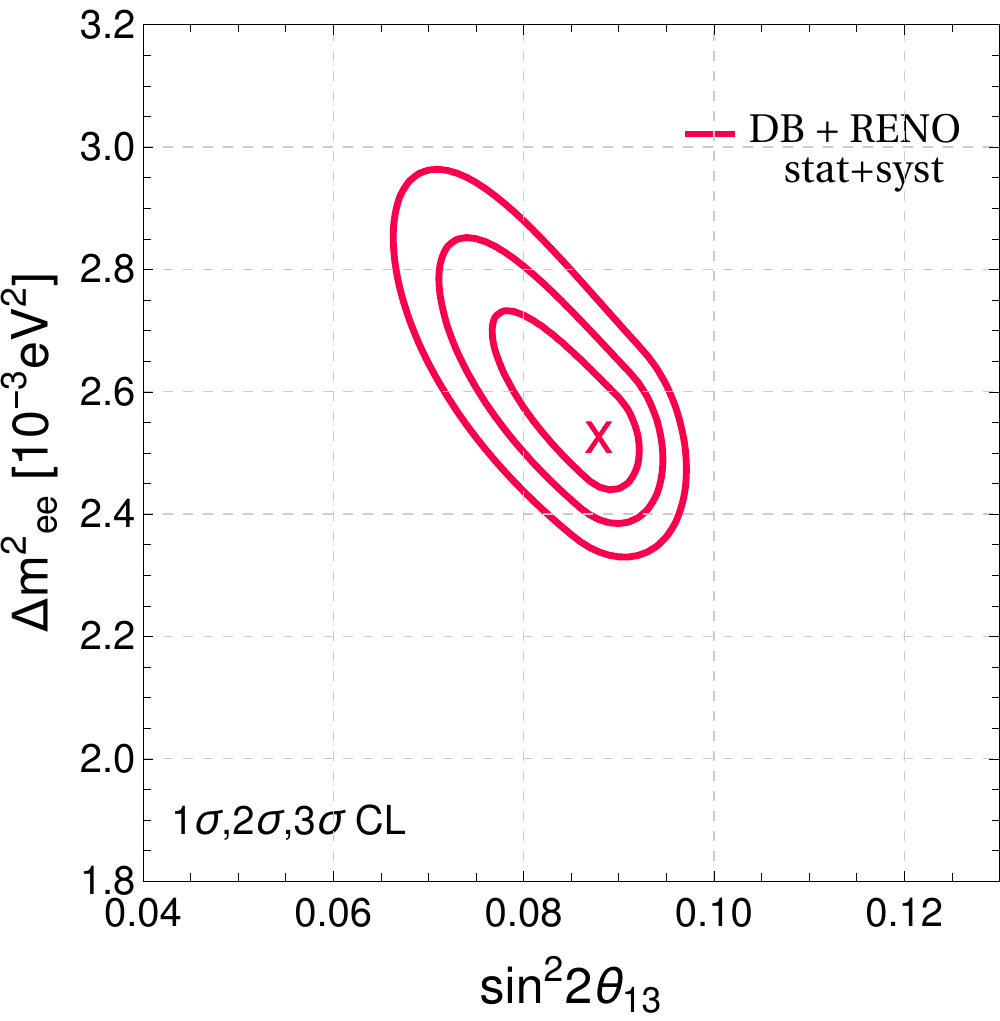}\hfill
\includegraphics[width=0.308\textwidth]{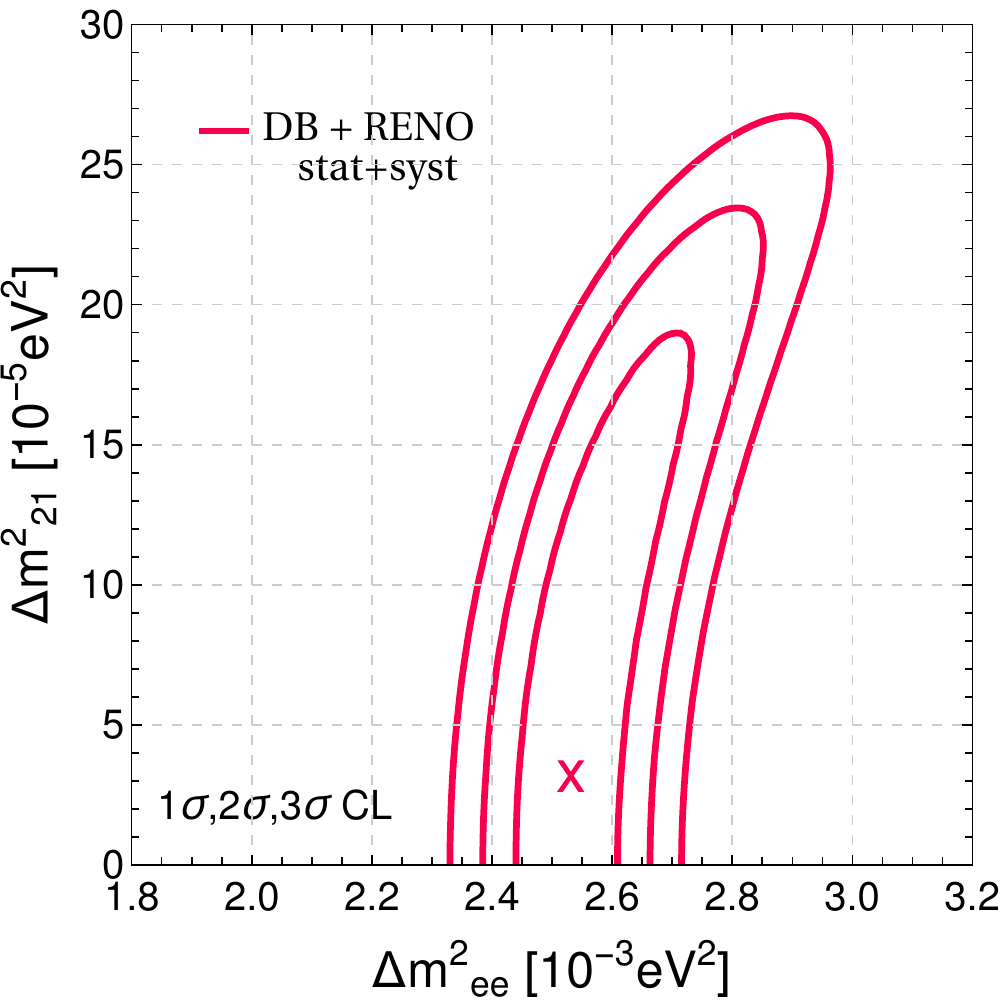}\\
\caption{Simultaneous three parameter fit for $\Delta m^2_{21}$,  $\Delta m^2_{ee}$ and $\sin^2 2 \theta_{13}$
using the combined Daya Bay (1,958 days, 487 K IBD events at Far) and RENO (2,200 days, 98 K IBD events at Far) data.
The best fit point is found at
$\Delta m^2_{21}=\,3.3 \times 10^{-5} {\rm eV}^2$,  $\Delta m^2_{ee} = \, 2.5 \times 10^{-3} {\rm eV}^2$,  $\sin^2 2 \theta_{13}  =\,  0.088 $.}
\label{fig:DB&RENO_three}
\end{figure*}

In Fig. \ref{fig:DB&RENO_three}, we give the constraints on the three parameter fit, $\Delta m^2_{21}$,  $\Delta m^2_{ee}$ and $\sin^2 2 \theta_{13}$,  without imposing any constrain on $\Delta m^2_{ee}$,
using the combined Daya Bay and RENO data sets. Both statistical and systematic uncertainties are included in this plot. As before $\theta_{12}$ is fixed at $\sin^2 \theta_{12} = 0.310$, see \cite{Seo:2018rrb} for discussion on allowing  $\sin 2 \theta_{12}$ to also vary.\\

Results with $\Delta m^2_{ee}$ fixed or free are obtained for each experiment and for when the data from both experiments are combined. 
These are described  and given in Appendix~\ref{a:fixedVsFree}. 
It was found that the effect of free $\Delta m^2_{ee}$ is bigger than that of systematic uncertainty, 
but our representing results are based on constrained $\Delta m^2_{ee}$ since it is a reasonably well measured oscillation parameter
using LBL experiments .

\section{\label{sec:summary} Conclusion}

Using the currently available public data from Daya Bay (1,958 days) and RENO (2,200 days), we have provided additional information 
on the solar $\Delta m^2$. 
A reasonable upper bound is obtained from a combined analysis of the Daya Bay and RENO data as 18 $\times 10^{-5}$ eV$^2$ at 95\% CL
, where $\Delta m^2_{ee}$ was constrained using a pull parameter with input information from LBL experiments.
Our combined analysis result is currently limited by statistics and, as expected, Daya Bay data drives the combined analysis results. 
Our analysis method was validated by reproducing the $\Delta m^2_{ee}$ and $\sin^2 \theta_{13}$ contours for each experiment as discussed in Appendix~\ref{a:validation}.
\\ 
Given that the previous measurements by KamLAND and SK/SNO of the solar $\Delta m^2$ are in a 2$\sigma$ tension and the importance of solar $\Delta m^2$ 
for the determination of CP violation in  LBL experiments, it is crucial that we understand the value of the solar $\Delta m^2$ better.
It is expected by circa 2025 that the JUNO experiment will provide additional, important information on the value of the of solar $\Delta m^2$.

\begin{acknowledgments}
We are grateful to Thomas Schwetz for fruitful discussions.
This work (SHS) was supported by the National Research Foundation of Korea (NRF) grant funded by the Korea Ministry of Science and ICT (MSIT)
(No. 2017R1A2B4012757 and IBS-R016-D1-2019-b01).
This manuscript has been authored (SJP) by Fermi Research Alliance, LLC under Contract No.~DE-AC02-07CH11359 with the U.S. Department of Energy, Office of Science, 
Office of High Energy Physics.
This project (SJP) has received funding/support from the European Union's Horizon 2020 research and innovation programme under the Marie Sklodowska-Curie grant 
agreement No 690575 \& No 674896.
\end{acknowledgments}



\begin{appendix}

\section{VALIDATION OF OUR ANALYSES}
\label{a:validation}
\begin{figure}[ht]
\begin{center}
\includegraphics[width=0.45\textwidth]{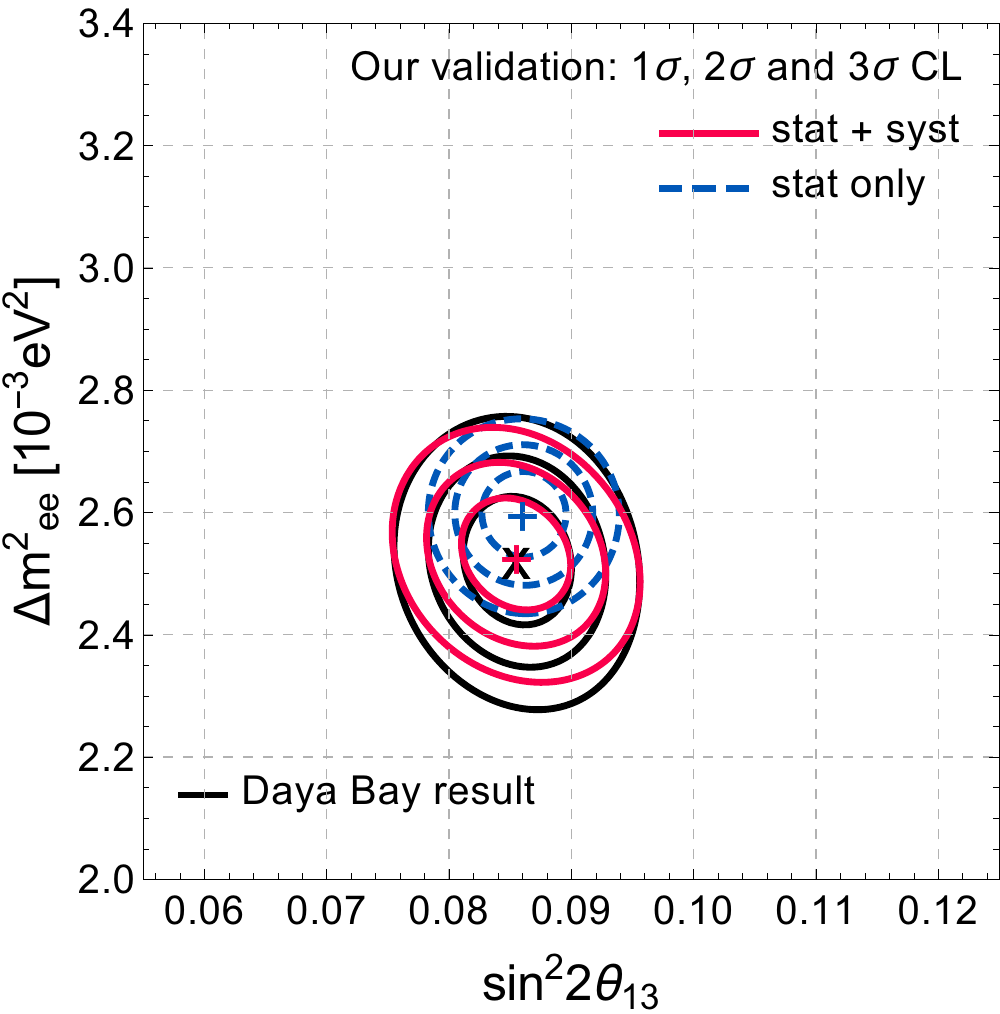}
\end{center}
\caption{Our validation on $\Delta m^2_{ee}$ vs $\sin^2 2 \theta_{13}$ fit using the Daya Bay data (1,958 days),
including systematics and statistics uncertainties in red solid lines,
and including statistics only in blue dashed lines, for 1, 2 and 3 $\sigma$ allowed regions. The fit of the Daya Bay collaboration with 1,958 days from
\cite{Adey:2018zwh} is represented in the solid black lines. 
The agreement between our analysis (solid red lines) and Daya Bay's analysis (solid black lines) is excellent.  }
\label{fig: DB dmsqee}
\end{figure}

\begin{figure}[ht]
\begin{center}
\includegraphics[width=0.45\textwidth]{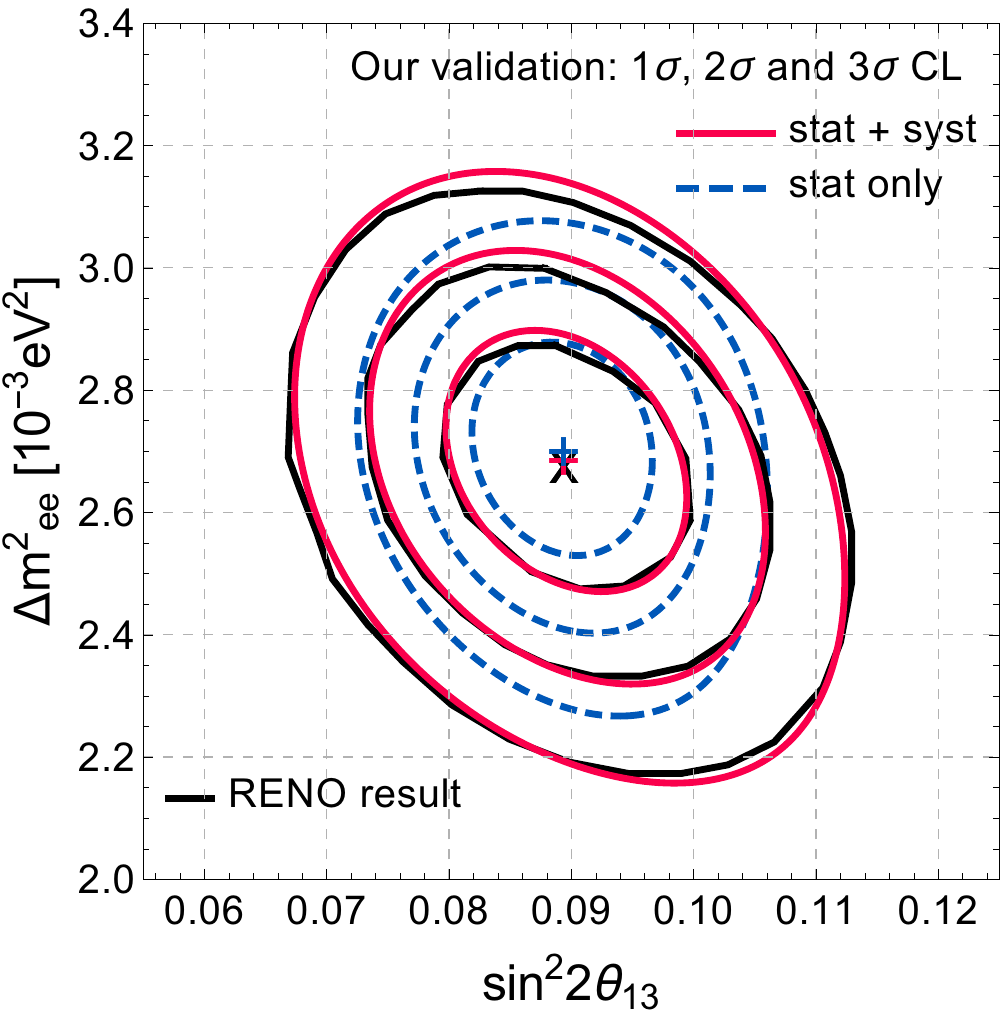}
\end{center}
\caption{Our validation on $\Delta m^2_{ee}$ vs $\sin^2 2 \theta_{13}$ fit using the RENO data (2,200 days),
including systematics and statistics uncertainties in red solid lines,
and including statistics only in blue dashed lines, for 1, 2 and 3 $\sigma$ allowed regions.  The fit of the RENO collaboration with 2,200 days
from \cite{Bak:2018ydk} is represented in the solid black lines. 
The agreement between our analysis (solid red lines) and RENO's analysis (solid black lines) is excellent.}
\label{fig: RENO dmsqee}
\end{figure}

Using the data and the $\chi^2$ formalism described in section III and IV, our method reproduces the contours in the $\Delta m^2_{ee}$ vs $\sin^2 2\theta_{13}$
from the the Daya Bay and RENO collaborations as it is shown in figures \ref{fig: DB dmsqee}, \ref{fig: RENO dmsqee}.
The Day Bay and RENO collaboration contours are taken from the supplementary material of \cite{Adey:2018zwh} and from FIG. 3 of \cite{Bak:2018ydk}, respectively.

The agreement between our results and Daya Bay as well as RENO for the measurements of $\Delta m^2_{ee}$ vs $\sin^2 2\theta_{13}$
is an excellent validation of the methods and numbers used in our analysis. Therefore, our constraint on $\Delta m^2_{21}$, using the publicly available data of 
Daya Bay and RENO, has a firm basis.

\section{NUMBER OF EXPECTED EVENTS AND PULL PARAMETERS IN $\chi^2$}
\label{a:predEvt}
The expected numbers of signal events in a detector $d$ in a prompt energy bin $i$, $X_i^d$, is computed as follows up to a common input
(e.g. reactor power, total number of protons) which cancels when taking ratios in the $\chi^2$ computation.

\begin{eqnarray}
X_i^d &=& \sum_r \sum_{\text{iso}} \frac{a^d}{L^2_{rd}} \int^{E^{rec}_{i+1}}_{E^{\text{rec}}_i} {\rm d} E^{\text{rec}} \int^{\infty}_{0} {\rm d} E_{\nu} ~ \sigma(E_{\nu}) f^{\text{iso}} \phi^{\text{iso}} (E_{\nu}) \nonumber \\
&\times& P^{rd}_{\bar \nu_e \rightarrow \bar \nu_e}(E_{\nu}) R(E^{\text{rec}},E_{\nu})
\label{EQ: number of events}
\end{eqnarray}
where, the indices $i$, $r$, $d$, and iso refers to the $i^{th}$ energy bin, $r^{th}$ reactor, $d^{th}$ detector, and a fissionable isotope
($^{235}\rm U$, $^{239}\rm Pu$, $^{238}\rm U$, or $^{241}\rm Pu$), respectively, and $a^d$ is the detector efficiency.
$L_{rd}$ is the baseline between the reactor $r$ and the detector $d$.
$E_{\nu}$ and $E^{\text{rec}}$ are the neutrino true energy and the reconstructed energy, both related by the detector response function
$R(E^{\text{rec}},E_{\nu})$.
The $\sigma(E_{\nu})$ is the IBD cross section computed performing the integral in ${\rm d}\! \cos \theta$ of the differential cross section in \cite{Vogel:1999zy}
and the $f^{\text{iso}}$ is the averaged fission fraction\footnote{Ideally we would have the information on the fission factions as a function of time in each reactor, 
but since we do not have this information we take the same averaged values for all the detectors. This means that any systematic uncertainty on the flux predictions 
will cancel when taking ratios of the expected events in different experimental sites.}
and the $\phi^{\text{iso}}(E_{\nu})$ is the Huber-Mueller flux prediction \cite{Huber:2011wv,Mueller:2011nm}.
$P^{rd}_{\bar \nu_e \rightarrow \bar \nu_e}(E_{\nu})$ is the oscillation probability from reactor r to detector d in the three neutrino oscillation paradigm.

\begin{figure*}[t]
\centering
\includegraphics[width=0.3\textwidth]{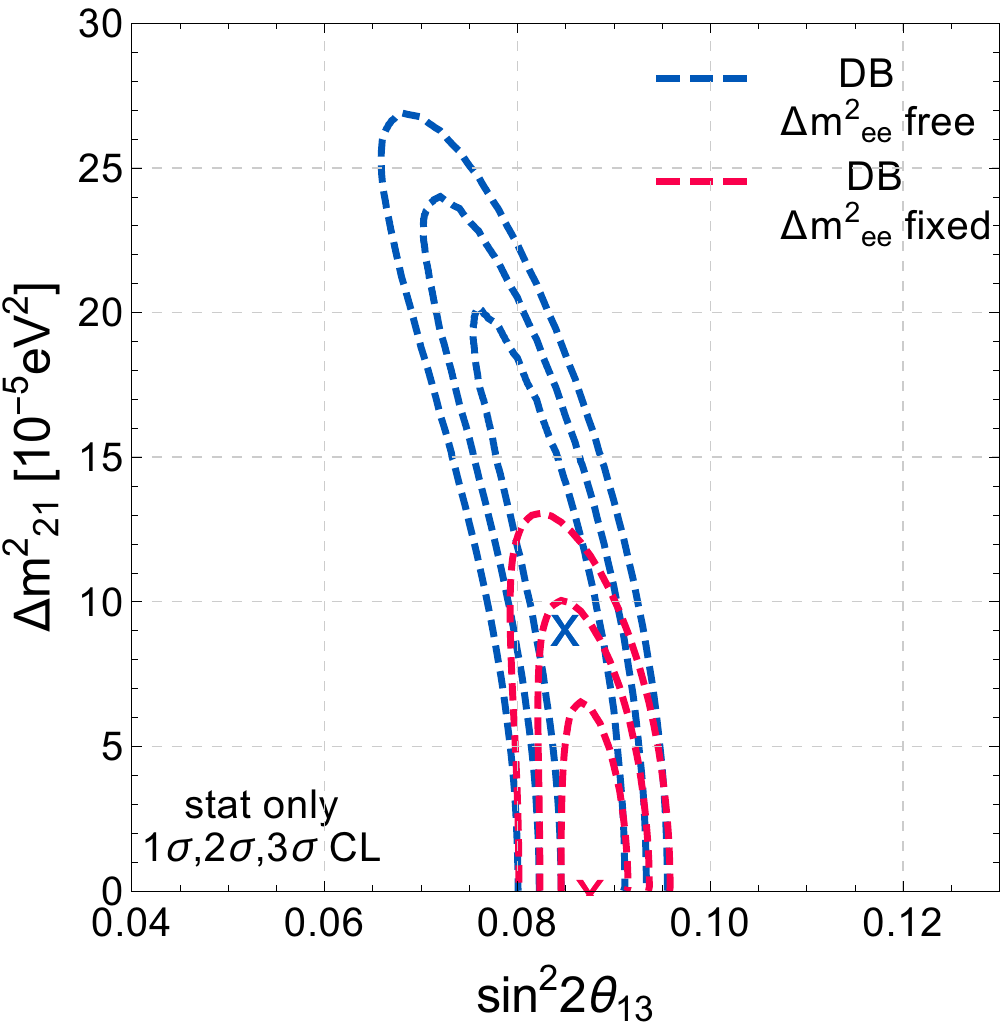}
\includegraphics[width=0.3\textwidth]{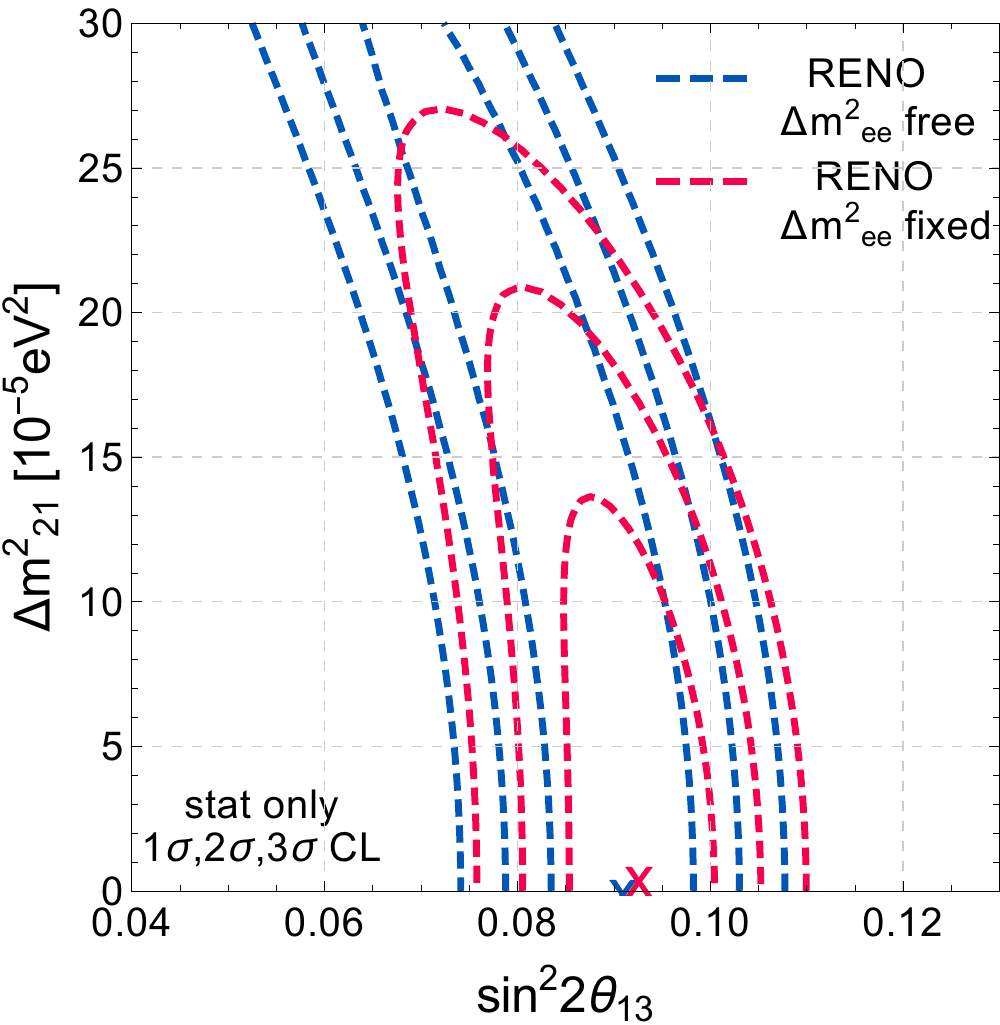}
\includegraphics[width=0.3\textwidth]{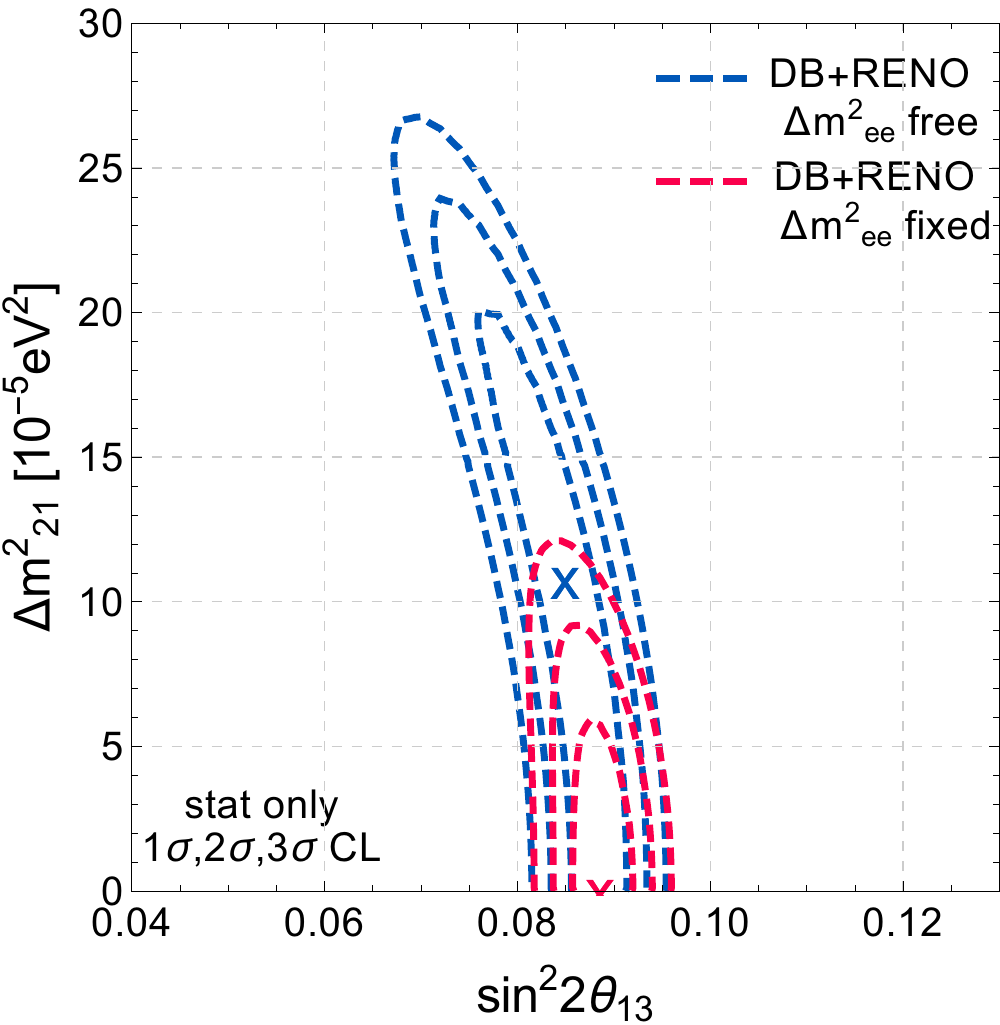}\\
\includegraphics[width=0.3\textwidth]{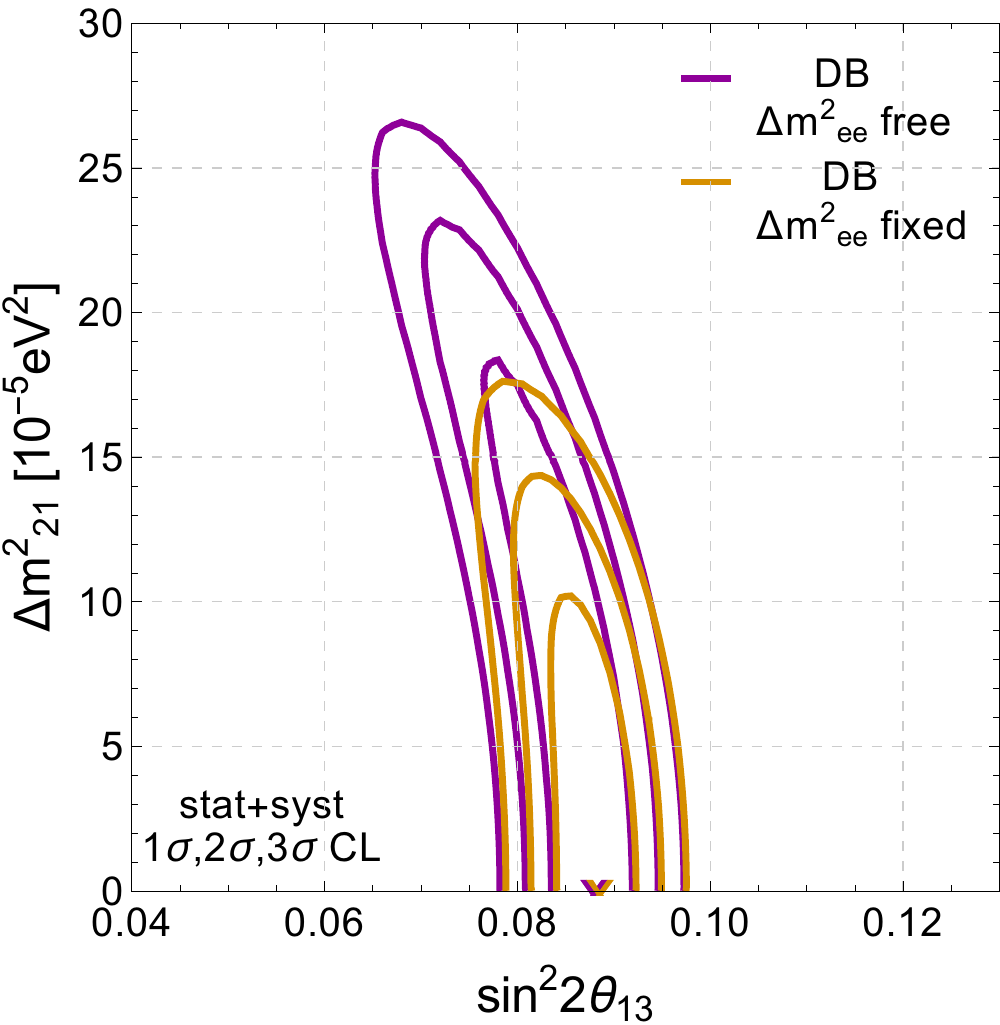}
\includegraphics[width=0.3\textwidth]{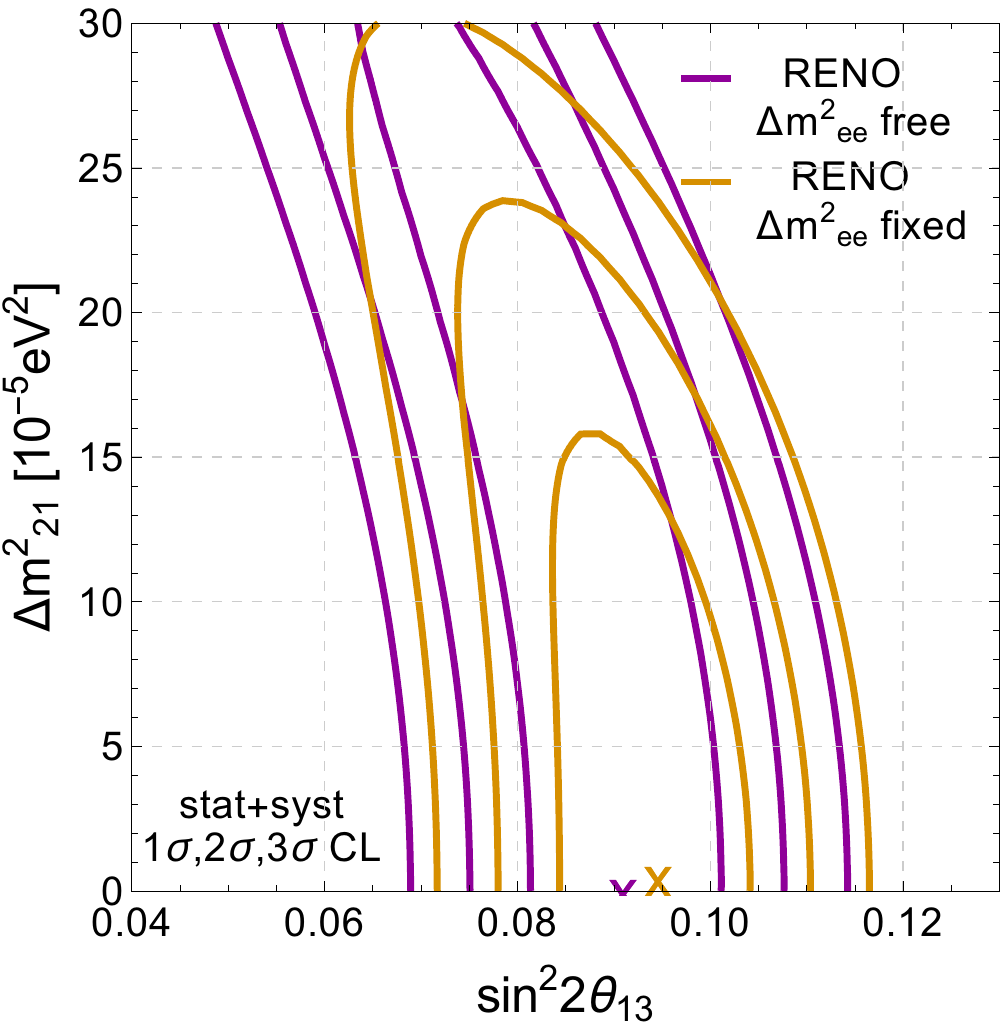}
\includegraphics[width=0.3\textwidth]{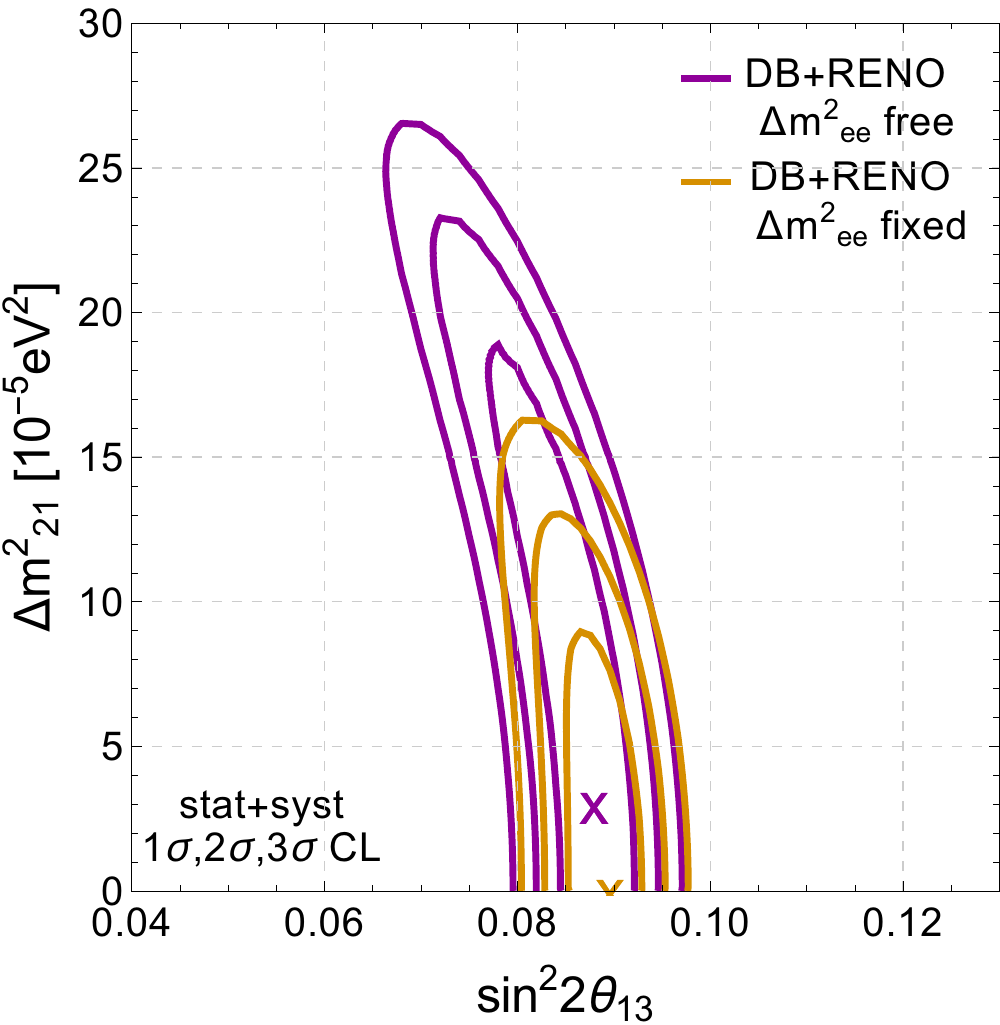}\\
\caption{Daya Bay (left panels), RENO (middle panels) and combined (right panels) 1, 2 and 3 $\sigma$ allowed regions in the $\Delta m^2_{21}$ vs $\sin^2 2 \theta_{13}$ 
parameter space, leaving $\Delta m^2_{ee}$ free or fixed at $2.45 \times 10^{-3} {\rm \, eV}^2$.
For all panels, the dashed lines include only statistical uncertainties whereas the solid lines include  both statistical and systematic uncertainties.
The best fit values are shown with $\times$ signs. This figure demonstrates the extremes of our result to variation of $\Delta m^2_{ee}$.  }
\label{fig:DB&RENO_free_fixed}
\end{figure*}

The pull parameters accounting for detection efficiency ($\epsilon^d$) and relative energy scale ($\eta^d$) are included in the number of expected events as follows
\begin{eqnarray*}
X_i^d(\epsilon^d, \eta^d) &=& \epsilon^d \sum_r \sum_{\text{iso}} \frac{a^d}{L^2_{rd}} \int^{\eta^d E^{rec}_{i+1}}_{\eta^d E^{\text{rec}}_i} {\rm d} E^{\text{rec}} \int^{\infty}_{0} {\rm d} E_{\nu} ~ \nonumber \\ &\times&
\sigma(E_{\nu}) f^{\text{iso}} \phi^{\text{iso}} (E_{\nu}) P^{rd}_{\bar \nu_e \rightarrow \bar \nu_e}(E_{\nu}) R(E^{\text{rec}},E_{\nu}) \, .
\end{eqnarray*}
For RENO, the efficiency pull parameter is included in the ratio.

The background pull parameters are included in background events $B_i^d$ used in $D^{F/N}_i \equiv \frac{O^F_i-B^F_i}{O^N_i-B^N_i}$ as follows
\begin{eqnarray*}
B_i^d(b^d_{\rm LH},b^d_{\rm acc},b^d_{\rm n}) &=& B_i^d + (b^d_{\rm LH}-1) B^d_{{\rm LH}, i} \nonumber \\
&+& (b^d_{\rm acc}-1) B^d_{{\rm acc}, i} + (b^d_{\rm n}-1) B^d_{{\rm n}, i} \, ,
\end{eqnarray*}
where $B_i^d$ ($B_{{\rm LH},i}^d$, $B_{{\rm acc}, i}^d$ and $B_{{\rm n}, i}^d$) represents the number of total (Li-He, accidental and fast neutron) background events
in the $i^{th}$ prompt energy bin in the $d^{th}$ detector,
and the small $b$ represents the corresponding pull parameter.

\section{FIXED VS FREE $\Delta m^2_{ee}$}
\label{a:fixedVsFree}
For the results in the main body of our paper we constrained $\Delta m^2_{ee}$ treating it as a pull parameter  using LBL experiments input. In this Appendix we show the impact of $\Delta m^2_{ee}$ 
fixed and set free.
A 2-dimensional scan over $\Delta m^2_{21}$ and $\sin^2 2 \theta_{13}$ is performed to find the best fit value pair at the minimum value of $\chi^{2}$
described earlier, where in the oscillation probability $\theta_{12}$ is fixed as $\sin^2 \theta_{12} = 0.310$ but $\Delta m^2_{ee}$ is set free within the range of
$ [1.55, 3.55] \times 10^{-3}$ eV$^{2}$.
Results with a fixed $\Delta m^2_{ee} = 2.45 \times 10^{-3} {\rm eV}^2$ are also obtained and compared to those with
$\Delta m^2_{ee}$ set free.
Figure~\ref{fig:DB&RENO_free_fixed}, left and middle panels, shows the results of $\Delta m^2_{ee}$ fixed and free for Daya Bay and RENO.
It is observed that the effect of floating $\Delta m^2_{ee}$ is bigger than adding systematic uncertainty for both Daya Bay and RENO.
For floating $\Delta m^2_{ee}$ case, the corresponding $\Delta m^2_{ee}$ values for the minimum $\chi^2$ are found to be
$2.50\times 10^{-3}$eV$^2$ ($2.68\times 10^{-3}$ eV$^2$) for Daya Bay (RENO) and it is within 1 $\sigma$ uncertainty of each of their measurements.
Figure~\ref{fig:DB&RENO_free_fixed}, right panels  shows the results with combined analysis.
For floating $\Delta m^2_{ee}$ case, the corresponding $\Delta m^2_{ee}$ value for the minimum $\chi^2$ is found to be
$2.54\times 10^{-3}$eV$^2$ and it is within 1 $\sigma$ uncertainty of the Daya Bay best fit value, i.e., $ [2.52 \pm 0.07]\times 10^{-3}$.\\

\end{appendix}


\begin{thebibliography}{99}

\bibitem{Seo:2018rrb} 
  S.~H.~Seo and S.~J.~Parke,
  Phys.\ Rev.\ D {\bf 99}, no. 3, 033012 (2019)
  doi:10.1103/PhysRevD.99.033012
  [arXiv:1808.09150 [hep-ex]].

\bibitem{An:2015rpe}
  F.~P.~An {\it et al.} [Daya Bay Collaboration],
  Phys.\ Rev.\ Lett.\  {\bf 115}, no. 11, 111802 (2015)
  doi:10.1103/PhysRevLett.115.111802
   {[arXiv:1505.03456v2 [hep-ex]]}.

\bibitem{RENO:2015ksa} 
  J.~H.~Choi {\it et al.} [RENO Collaboration],
  Phys.\ Rev.\ Lett.\  {\bf 116}, no. 21, 211801 (2016)
  doi:10.1103/PhysRevLett.116.211801
  [arXiv:1511.05849 [hep-ex]].


\bibitem{Abe:2010hy}
  K.~Abe {\it et al.} [Super-Kamiokande Collaboration],
  Phys.\ Rev.\ D {\bf 83}, 052010 (2011)
  doi:10.1103/PhysRevD.83.052010
  [arXiv:1010.0118 [hep-ex]].
  

\bibitem{Aharmim:2011vm}
  B.~Aharmim {\it et al.} [SNO Collaboration],
  Phys.\ Rev.\ C {\bf 88}, 025501 (2013)
  doi:10.1103/PhysRevC.88.025501
  [arXiv:1109.0763 [nucl-ex]].
  
  
\bibitem{Gando:2010aa}
  A.~Gando {\it et al.} [KamLAND Collaboration],
  Phys.\ Rev.\ D {\bf 83}, 052002 (2011)
  doi:10.1103/PhysRevD.83.052002
  [arXiv:1009.4771 [hep-ex]].

\bibitem{An:2015jdp}
  F.~An {\it et al.} [JUNO Collaboration],
  J.\ Phys.\ G {\bf 43}, no. 3, 030401 (2016)
  doi:10.1088/0954-3899/43/3/030401
  [arXiv:1507.05613 [physics.ins-det]].

\bibitem{Ayres:2004js}
  D.~S.~Ayres {\it et al.} [NOvA Collaboration],
  hep-ex/0503053.

\bibitem{Abe:2011ks} 
  K.~Abe {\it et al.} [T2K Collaboration],
  Nucl.\ Instrum.\ Meth.\ A {\bf 659}, 106 (2011)
  doi:10.1016/j.nima.2011.06.067
  [arXiv:1106.1238 [physics.ins-det]].


\bibitem{Esteban:2018azc} 
  I.~Esteban, M.~C.~Gonzalez-Garcia, A.~Hernandez-Cabezudo, M.~Maltoni and T.~Schwetz,
  JHEP {\bf 1901}, 106 (2019)
  doi:10.1007/JHEP01(2019)106
  [arXiv:1811.05487 [hep-ph]].

\bibitem{Acciarri:2015uup}
  R.~Acciarri {\it et al.} [DUNE Collaboration],
  arXiv:1512.06148 [physics.ins-det].
  
  
\bibitem{Abe:2015zbg} 
  K.~Abe {\it et al.} [Hyper-Kamiokande Proto- Collaboration],
  PTEP {\bf 2015}, 053C02 (2015)
  doi:10.1093/ptep/ptv061
  [arXiv:1502.05199 [hep-ex]].

\bibitem{Abe:2016ero} 
  K.~Abe {\it et al.} [Hyper-Kamiokande Collaboration],
  PTEP {\bf 2018}, no. 6, 063C01 (2018)
  doi:10.1093/ptep/pty044
  [arXiv:1611.06118 [hep-ex]].


\bibitem{Abe:2017vif} 
  K.~Abe {\it et al.} [T2K Collaboration],
  Phys.\ Rev.\ D {\bf 96}, no. 9, 092006 (2017)
  Erratum: [Phys.\ Rev.\ D {\bf 98}, no. 1, 019902 (2018)]
  doi:10.1103/PhysRevD.96.092006, 10.1103/PhysRevD.98.019902
  [arXiv:1707.01048 [hep-ex]].

 
\bibitem{Abe:2018uyc} 
  K.~Abe {\it et al.} [Hyper-Kamiokande Collaboration],
  arXiv:1805.04163 [physics.ins-det].

 \bibitem{Beacom:2018xyz}
 F.~Capozzi, S.~W.~Li, G.~Zhu, J.~F.~Beacom,
 ``DUNE as the Next-Generation Solar Neutrino Experiment,''
 [1808.08232[hep-ph]].
  
\bibitem{Nunokawa:2005nx}
  H.~Nunokawa, S.~J.~Parke and R.~Zukanovich Funchal,\\
  Phys.\ Rev.\ D {\bf 72}, 013009 (2005),
  {[hep-ph/0503283]} 

\bibitem{Parke:2016joa}
  S.~Parke,
  Phys.\ Rev.\ D {\bf 93}, no. 5, 053008 (2016)
  doi:10.1103/PhysRevD.93.053008
  [arXiv:1601.07464 [hep-ph]].
  
  
  
\bibitem{Adey:2018zwh} 
  D.~Adey {\it et al.} [Daya Bay Collaboration],
  Phys.\ Rev.\ Lett.\  {\bf 121}, no. 24, 241805 (2018)
  doi:10.1103/PhysRevLett.121.241805
  [arXiv:1809.02261 [hep-ex]].
  
\bibitem{Bak:2018ydk} 
  G.~Bak {\it et al.} [RENO Collaboration],
  Phys.\ Rev.\ Lett.\  {\bf 121}, no. 20, 201801 (2018)
  doi:10.1103/PhysRevLett.121.201801
  [arXiv:1806.00248 [hep-ex]].
\bibitem{Seo:2014xei}
  S.~H.~Seo [RENO Collaboration].
  AIP Conf.\ Proc.\ {\bf 1666}, 080002 (2015)
  doi:10.1063/1.4915563
  [arXiv:1410.7987 [hep-ex]].

\bibitem{Vogel:1999zy} 
  P.~Vogel and J.~F.~Beacom,
  Phys.\ Rev.\ D {\bf 60}, 053003 (1999)
  doi:10.1103/PhysRevD.60.053003
  [hep-ph/9903554].

\bibitem{Huber:2011wv} 
  P.~Huber,
  Phys.\ Rev.\ C {\bf 84}, 024617 (2011)
  Erratum: [Phys.\ Rev.\ C {\bf 85}, 029901 (2012)]
  doi:10.1103/PhysRevC.85.029901, 10.1103/PhysRevC.84.024617
  [arXiv:1106.0687 [hep-ph]].
  
\bibitem{Mueller:2011nm} 
  T.~A.~Mueller {\it et al.},
  Phys.\ Rev.\ C {\bf 83}, 054615 (2011)
  doi:10.1103/PhysRevC.83.054615
  [arXiv:1101.2663 [hep-ex]].

\end{thebibliography}
\end{document}